\documentclass[useAMS,usenatbib]{mn2e}

\usepackage{aecompl}
\usepackage[toc,page]{appendix}

\usepackage{graphicx}
\usepackage{amssymb}
\usepackage{amsmath}
\usepackage{aas_macros}
\usepackage{deluxetable}
\usepackage{lscape}
\usepackage{rotating}
\usepackage[symbol*]{footmisc}
\usepackage{gensymb}
\usepackage{tikz}



\title[RJ-plots]{RJ-plots: An improved method to classify structures objectively}

\author[S. D. Clarke et al.]{S. D. Clarke$^{1}$\thanks{E-mail: sclarke@asiaa.sinica.edu.tw }, S. E. Jaffa$^{2}$ and A. P. Whitworth$^{3}$\\$^{1}$Academia Sinica, Institute of Astronomy and Astrophysics, Taipei, Taiwan\\$^{2}$UCL Advanced Research Computing Centre, Gower Street, London WC1E 6BT, UK\\$^{3}$School of Physics and Astronomy, Cardiff University, Cardiff, CF24 3AA, UK}

\newcommand{\Su}{_{_{\odot}}}
\begin{document}

\date{}

\pagerange{\pageref{firstpage}--\pageref{lastpage}} \pubyear{2002}

\maketitle

\label{firstpage}

\begin{abstract}
The interstellar medium is highly structured, presenting a range of morphologies across spatial scales. The large data sets resulting from observational surveys and state-of-the-art simulations studying these hierarchical structures means that identification and classification must be done in an automated fashion to be efficient. Here we present RJ-plots, an improved version of the automated morphological classification technique J-plots developed by \citet{Jaf18}. This method allows clear distinctions between quasi-circular/elongated structures and centrally over/under-dense structures. We use the recent morphological SEDIGISM catalogue of \citet{Ner22} to show the improvement in classification resulting from RJ-plots, especially for ring-like and concentrated cloud types. We also find a strong correlation between the central concentration of a structure and its star formation efficiency and dense gas fraction, as well as a lack of correlation with elongation. Furthermore, we use the accreting filament simulations of \citet{Cla20} to highlight a multi-scale application of RJ-plots, finding that while spherical structures become more common at smaller scales they are never the dominant structure down to $r\sim0.03$ pc.

\end{abstract}


\section{Introduction}\label{SEC:INTRO}%
The interplay between gravity, magnetic fields, turbulence, radiation transport, chemistry and stellar feedback shapes the interstellar medium (ISM) into a complex web of structures with varying morphologies across multiple scales. Large-scale observational survey data have shown the prevalence of bubbles and ring-like structures associated with HII regions and super-shells seen in HI linked to multiple supernovae, as well as the connection between dense gas filaments and cores with the star formation process \citep{McC02,Chu06,Dai07,Mol10,And10,Ehl13,Sch14,Li16,Rig16,Arz19,Sch20,Col21,Ner22}. It is imperative for methods to readily identify and classify these features as it is only through understanding the formation of, connection between, and gas flows within these structures that a comprehensive picture of the ISM can be built \citep[see][for a recent review on the topic]{Pin22}.

The large volume of data from observational surveys, as well as state-of-the-art simulations, has made the automated identification and classification of structures a necessity. Multiple algorithms have been developed to identify specific types of structures, such as filaments \citep[e.g.][]{Sou11,Sch14,Men21} and quasi-circular cores/clumps \citep[e.g.][]{Stu90,Men12,Ber15} so that they may be studied. Another approach is to identify structures not by some desired geometry but by selecting contiguous areas/volumes of an image/cube which shares similar properties, such as column densities or line emission \citep[e.g. dendrograms and SCIMES,][]{Ros08,Col15}. In these cases, there is no pre-selection for structure morphology and it must be determined in a post-hoc manner. In this paper we focus on this post-hoc morphology classification and confine ourselves to 2D images of structures (e.g. column density maps or moment zero maps of structures identified in position-position-velocity space.)

\subsection{J-plots}\label{SSEC:JPLOTS}%
An automated technique to identify and classify the morphology of 2D structures called J-plots is presented in \citet{Jaf18}. This technique constructs two quantities, called J-values, by first calculating the principal moments of inertia of a structure, I$_1$ and I$_2$, and comparing it to an equivalent circle of equal area and total weight\footnote{\citet{Jaf18} refer to this quantity as the mass of the structure as they are concerned with structures identified in column density images. We adopt total weight due to its generality.}, $a$ and $W$. The total weight of a structure is the integral of the weighting of each pixel (e.g. column density or intensity); the exact quantity is unimportant. The equivalent circle has a moment of inertia of:
\begin{equation}
\mathrm{I_0} = \frac{aW}{4\pi}.
\end{equation}
The J-values may thus be defined as:
\begin{equation}
\mathrm{J}_i = \frac{\mathrm{I_0 - I}_i}{\mathrm{I_0 + I}_i}, \;\;\;\; i = 1,2.
\end{equation}
By construction J$_2$ $\leq$ J$_1$. These two quantities are plotted against each other to produce a J-plot which may be used for classification (an example is the left panel of figure \ref{fig::RJplot}). A structure which has uniform surface density and is perfectly circular has J$_1$=J$_2$=0 and lies at the origin of a J-plot; a centrally concentrated quasi-circular structure has J$_1$ $>$ 0, J$_2$ $>$ 0 and lies in the upper-right quadrant of a J-plot (a J-core); a centrally under-dense quasi-circular structure has J$_1$ $<$ 0, J$_2$ $<$ 0 and lies in the bottom-left quadrant of a J-plot (a J-bubble); and an elongated structure has J$_1$ $>$ 0, J$_2$ $<$ 0 and lies in the bottom-right quadrant of a J-plot (a J-filament). 

This method provides a simple and efficient method for the classification of structures in simulations, real observations and synthetic observations \citep{Cla18,Pet21,Ner22}. However, due to the construction that J$_2$ $\leq$ J$_1$, only half of the J-space may be occupied; importantly while the entire bottom-right quadrant (elongated structures) may be filled, only half of the upper-right and bottom-left (centrally concentrated and rarefied structures respectively) may be. This leads to a bias towards identifying structures as filaments and against identifying ring-like or concentrated structures. 

To help compensate for the above-mentioned biases, we develop and present an improved version of J-plots called RJ-plots. This paper is structured in the following manner: in section \ref{SEC:RJ} we describe the improvements made to J-plots as well as test and calibrate the RJ-plot technique against a number of basic structure shapes; in section \ref{SEC:APP} we show a selection of applications of RJ-plots using both observational data (section \ref{SSEC:SED}) and simulation data (section \ref{SSEC:SIM}); and in section \ref{SEC:CON} we conclude.

\begin{figure*}
\centering
\includegraphics[width=0.99\linewidth]{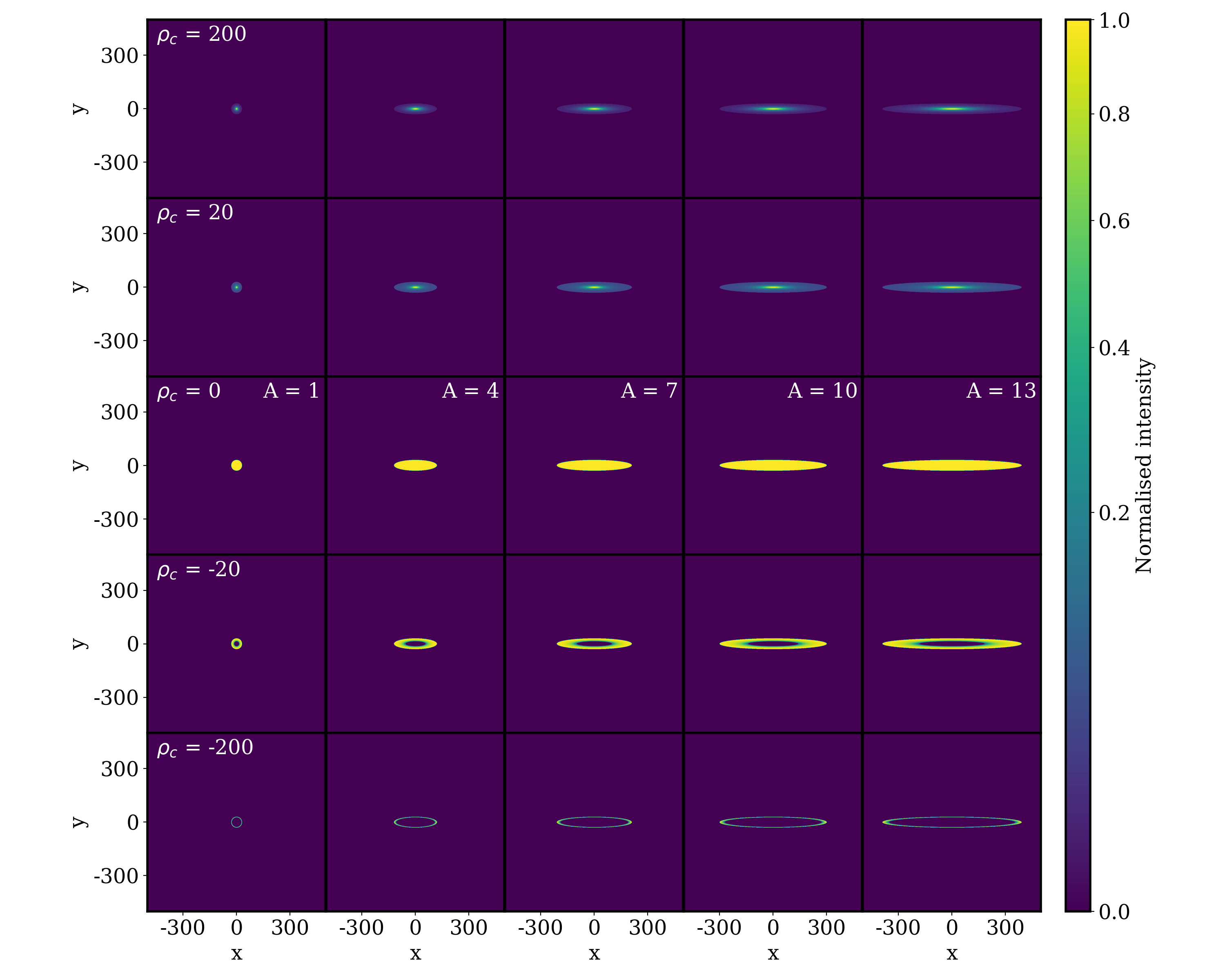}
\caption{A plot showing a selection of shapes parametrised in the manner described in section \ref{SSEC:BASIC}. The columns going from left to right show increasing aspect ratio, while rows going from top to bottom show a change from central over-densities to central under-densities. The parameters for the rows and columns are seen as white text in the figure. All shapes have been normalized such that their peak value is equal to 1.}
\label{fig::shapes}
\end{figure*}  

\section{RJ-plots}\label{SEC:RJ}%

\subsection{Basic shapes}\label{SSEC:BASIC}%
To demonstrate how the characteristics of a structure influence its resulting J/RJ-values, it is constructive to use a set of readily parametrisable basic shapes. Here we use ellipsoidal Plummer-like density profiles to produce a range of shapes parametrized using their aspect ratio, $A$, and their central density $\rho_c$.  Examples of these shapes can be seen in figure \ref{fig::shapes}. 

When constructing the shapes we use 1000x1000 pixel images to minimise grid artefacts from attempting to capture curved shapes with a Cartesian grid. The surface density of a shape is given as:
\begin{multline}
\rho(x,y) = \max{\left(\frac{\rho_c}{(1 + (x/x_0)^2 + (y/y_0)^2 )^2} + 1 \; | \; 0 \right)}\\\mathrm{if}  (x/x_0)^2 + (y/y_0)^2 < 16;
\end{multline}
otherwise the density is set to zero. Here the aspect ratio A is define as $x_0/y_0$ and $y_0$ is set to 8 pixels, allowing aspect ratios up to 15 to be considered. One can readily see that increasing $\rho_c$ produces a centrally over-dense structure while setting $\rho_c$ to a negative value produces a centrally under-dense structure. Taking the maximum between the Plummer-like profile and zero avoids negative surface densities when $\rho_c$ $<$ 0. These shapes are not to be considered necessarily typical shapes identified in the ISM, but are readily parametrisable and understood, allowing for a method to investigate the sensitivities of using moment of inertia based classification schemes. 

\begin{figure*}
\centering
\includegraphics[width=0.79\linewidth]{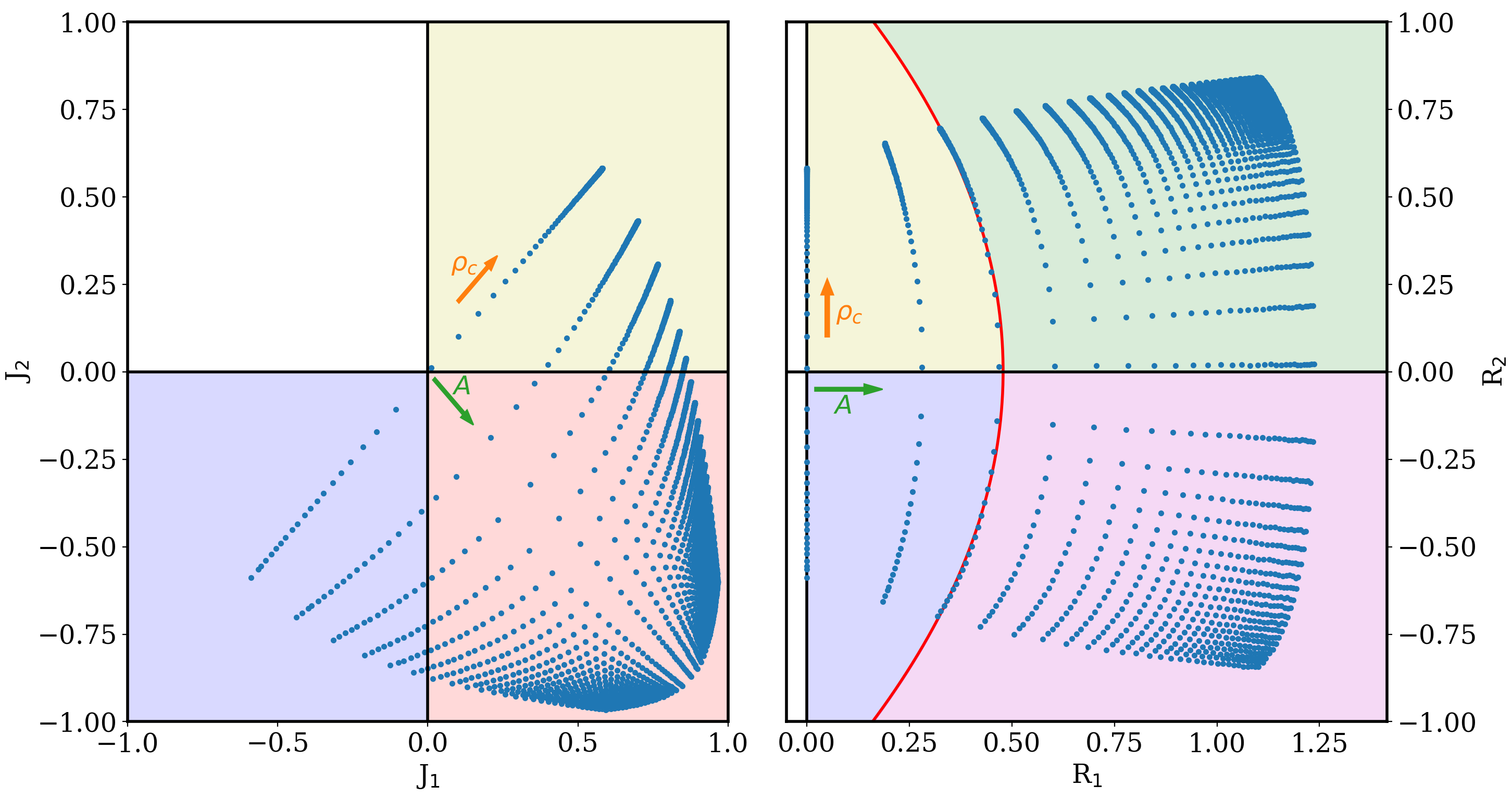}
\caption{(Left) A J-plot resulting from the shapes considered in section \ref{SSEC:BASIC}; and (right) the equivalent RJ-plot. The shapes are constructed with 1 $\leq$ A $\leq$ 15 in steps of 0.5, and with -100 $\leq$ $\rho_c$ $\leq$ 900 in steps of 5. In the J-plot, the quadrants are coloured blue, red and yellow to denote the classification of bubbles, filaments and cores respectively. In the RJ-plot, the A=2 shapes are fitted with a parabola to produce the red line, giving rise to 4 sections. The yellow and blue sections correspond to centrally over- and under-dense quasi-circular shapes; the green and magenta sections correspond to centrally over- and under-dense elongated shapes. The orange and green arrows show the approximate direction in which a structure moves with increasing $\rho_c$ and A respectively. }
\label{fig::RJplot}
\end{figure*}

The left panel of figure \ref{fig::RJplot} shows the J-plot resulting from considering shapes with 1 $\leq$ A $\leq$ 15 in steps of 0.5, and with -100 $\leq$ $\rho_c$ $\leq$ 900 in steps of 5. Each `track' in the J-plot is the result of a shape with constant aspect ratio and varying central density. The results from shapes with A=1 lie along the J$_2$=J$_1$ line, with centrally over-dense structures ($\rho_c$ $>$ 0) in the upper-right quadrant and centrally under-dense structure ($\rho_c$ $<$ 0) in the bottom left quadrant. Increasing aspect ratio leads to shapes moving into the bottom-right quadrant, with the $\rho_c$=0 shapes moving approximately along the J$_2$=-J$_1$ line.

By considering these basic shapes the biases in J-plots are apparent. Even small amounts of elongation moves structures quickly into the bottom-right elongated structure quadrant, with an A=1.5 and $\rho_c$=0 structure having (J$_1$,J$_2$) $\sim$ (0.25,-0.25), leading to a strong preference to classify objects as filaments. Furthermore, with increasing aspect ratio it becomes increasingly difficult for structures which are centrally over/under-dense to be correctly classified as they move more deeply into the elongated structure quadrant. Moreover, as $|$J$_i|$ $\leq$ 1, with increasing aspect ratio the available J-space to express differences in central density becomes increasingly constrained.

\begin{figure}
\centering
\includegraphics[width=0.95\linewidth]{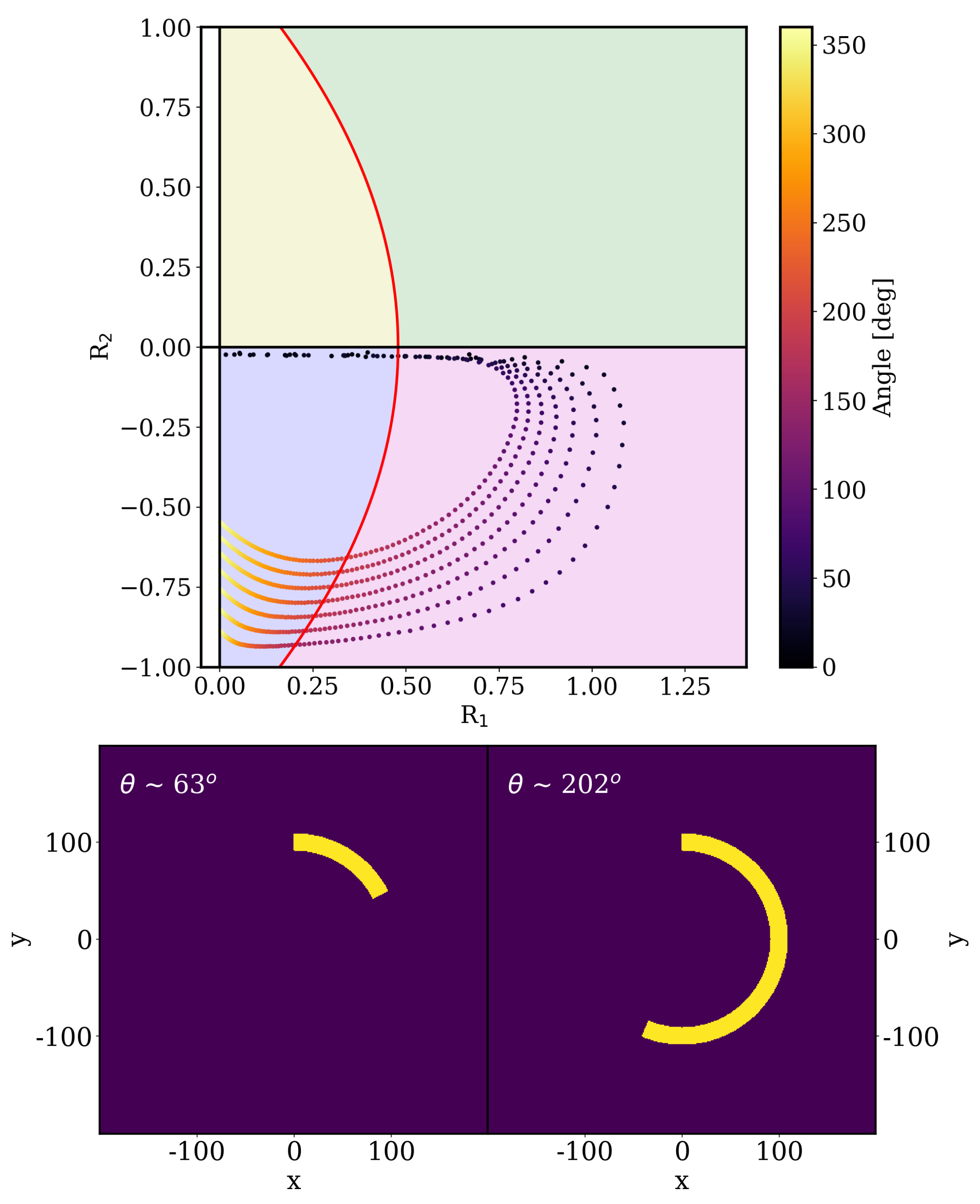}
\caption{(Top) An RJ-plot resulting from considering sectors of an annulus as a function of sector angle. The 7 tracks correspond to arcs with thicknesses between 6 and 30 pixels in steps of 4. The radius of each annuli is 100 pixels. (Bottom) The sector with the angle corresponding to the maximum R$_1$ value (left), and the sector with the angle where R$_1$, R$_2$ meet the red boundary line (right).}
\label{fig::rings}
\end{figure}  

\subsection{Rotated J-values}\label{SSEC:ROT}%
To address these limitations we propose that a more appropriate method for structure classification is to use rotated J-values, or RJ-values for short. As the two lines J$_2$=J$_1$ and J$_2$=-J$_1$ are seen to be important with shapes moving along them as their elongation and central density is modified, a rotation of 45 degrees anti-clockwise may be performed such that these lines become two orthogonal axes, i.e:
\begin{equation}
\mathrm{R}_1 = \frac{\mathrm{J_1 - J_2}}{\sqrt{2}} \; \; ; \; \; \mathrm{R}_2 = \frac{\mathrm{J_1 + J_2}}{\sqrt{2}}.
\end{equation}
With this step the two sets of information are quasi-separated with R$_1$ encoding elongation and R$_2$ encoding central under/over-density. Due to $|$J$_i|$ $\leq$ 1 there exist a maximum value of R$_2$ as a function R$_1$, i.e. R$_2$ = $\sqrt{2} -$ R$_1$, reducing the range of space to express differences in central density as elongation increases. This may be removed by normalising R$_2$ by this maximum value such that:
\begin{equation}
\mathrm{R}_2 = \frac{\mathrm{J_1 + J_2}}{\sqrt{2}(\sqrt{2} - \mathrm{R}_1)} = \frac{\mathrm{J_1 + J_2}}{2 - \mathrm{J_2 - J_1}}.
\end{equation}
This results in a space, 0 $\leq$ R$_1$ $\leq$ $\sqrt{2}$ and -1 $\leq$ R$_1$ $\leq$ 1, in which any point may be occupied by a shape. Being rotated and re-scaled values of J-values, RJ-values do not contain additional information about a shape's morphology but, as shown below, provide more explicit and easy to interpret results compared to J-values.

The RJ-plot resulting from considering the shapes described in section \ref{SSEC:BASIC} can be seen in the right-panel of figure \ref{fig::RJplot}. One sees that each track resulting from shapes with a given aspect ratio leads to a parabola; thus the two RJ-values do not completely separate the elongation information from the central density. However, the parabolae are shallow, showing a near separation of information, and that there is considerable improvement compared to J-values. Note also that neither axis responds linearly to changes in aspect ratio or central density, but they do respond monotonically.

It is important to discuss how best to interpret the value of R$_2$. While the basic shapes considered have either a single high-density region at their centre or a completely hollow, enclosed interior, this dichotomy is not what R$_2$ is measuring.  Rather, this quantity measures how weight is distributed with respect to the centre of weight of the object. Shapes where weight is predominately close to but not necessarily on top of the centre of weight of the object, e.g. a structure with multiple high-density clumps close to but not at its centre, will have high R$_2$ values, while structures where weight is predominately located far from the centre of weight, e.g. a highly curved filament where the centre of weight lies outside of the structure, will have low R$_2$ values. This point can be seen in figure \ref{fig::rings} which shows how sectors of an annulus move in RJ-space as the angle of the sector varies. When the angle is low, the centre of weight remains inside of the sector and R$_2$ $\sim$ 0. As the angle increases the sector appears increasingly filamentary as R$_1$ increases until at $\theta$ $\sim$ 60$\degree$ the centre of weight begins to move outside of the structure and R$_2$ decreases. As the angle continues to increase, the centre of weight shifts until it reaches approximately (0,0), at which point R$_2$ reaches its minimum and R$_1$ begins to decrease. When the annulus is complete it reaches R$_1$=0 and an R$_2$ value determined by its thickness. 

\begin{figure*}
\centering
\includegraphics[width=0.79\linewidth]{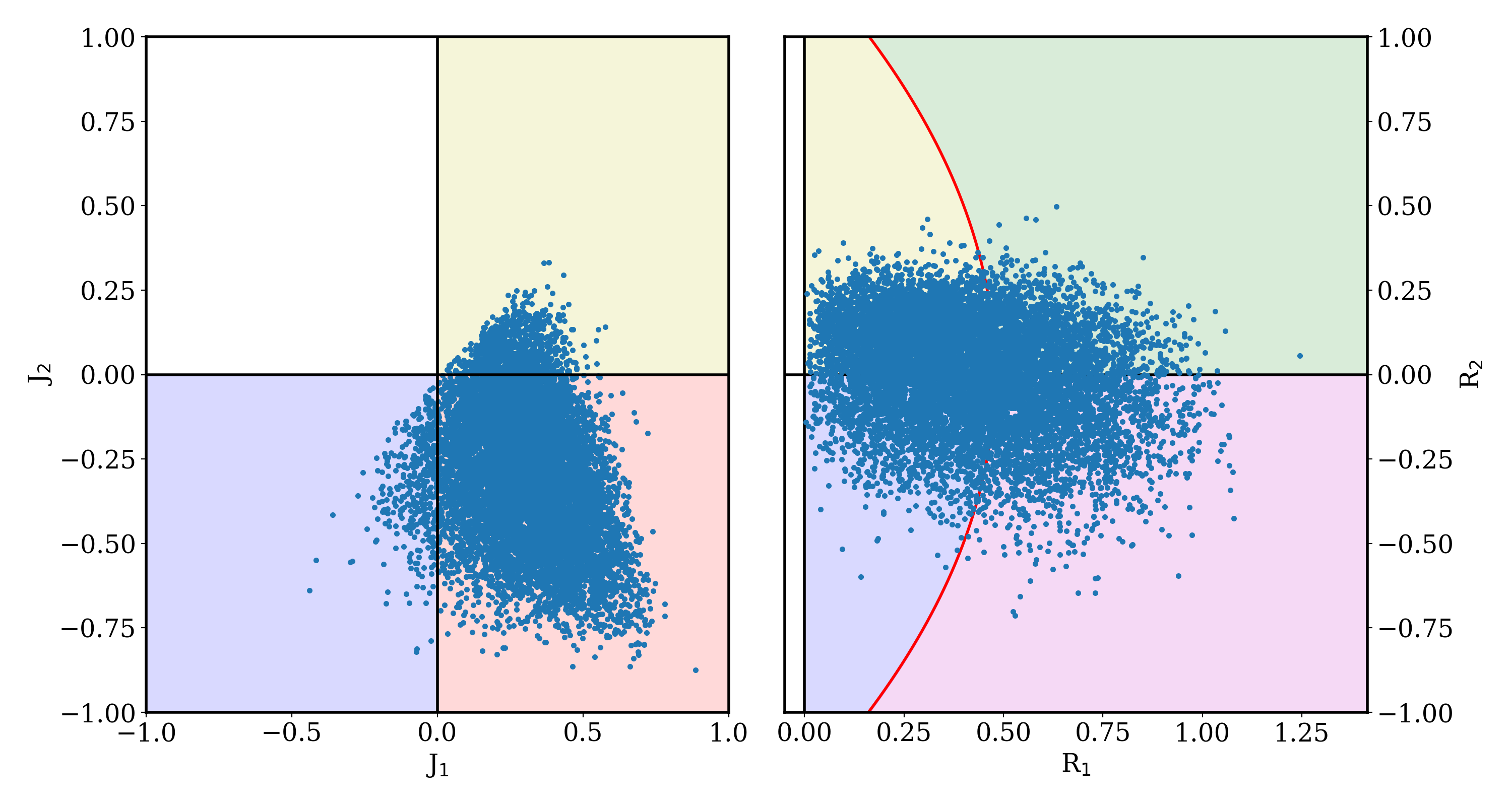}
\caption{(Left) A J-plot resulting from consider the 10,663 clouds identified by \citet{Ner22}, and (right) the equivalent RJ-plot. In the J-plot, the quadrants are coloured blue, red and yellow to denote the classification of bubbles, filaments and cores respectively. In the RJ-plot, the yellow and blue sections correspond to centrally over- and under-dense spherical shapes; and the green and magenta sections correspond to centrally over- and under-dense elongated shapes.}
\label{fig::Cat_RJ}
\end{figure*}  

Considering the discrete classification of structures using the RJ-plot, one can readily split the space into four sections. To separate between elongated and quasi-circular structures we use the parabola corresponding to the A=2 shapes; the exact choice of aspect ratio is arbitrary as any value greater than 1 is by definition elongated. The resulting fit yields the equation:
\begin{equation}
\mathrm{R}_1^E = 0.479 + 0.001 \mathrm{R}_2 - 0.317 \mathrm{R}_2^2,
\end{equation}
which can be seen as the red line in figure \ref{fig::RJplot}. Thus, a structure where R$_1$ $>$ R$_1^E$ is regarded as elongated and regarded as quasi-circular when the inverse is true. Further, the R$_2$=0 line may be used to separate structures which are centrally over-dense (R$_2$ $>$ 0) from those which are centrally under-dense (R$_2$ $<$ 0). Note that as shown in figure \ref{fig::rings}, structures with R$_1$ $>$ R$_1^E$ and R$_2$ $<$ 0, are most likely curved filaments and partial shells rather than strongly elongated bubbles, while half-to-complete shells lie in the region R$_1$ $<$ R$_1^E$ and R$_2$ $<$ 0.

\section{Applications}\label{SEC:APP}%
Here we will highlight two applications of RJ-plots: studying the morphology of clouds identified in the SEDIGISM survey and relating their morphology to cloud properties, and using numerical simulation data to investigate the morphology of structures formed on multiple scales in a filament. The application of RJ-plots is not limited to these two approaches but they do represent two types of study which may be productive in future work.

\begin{figure*}
\centering
\includegraphics[width=0.99\linewidth]{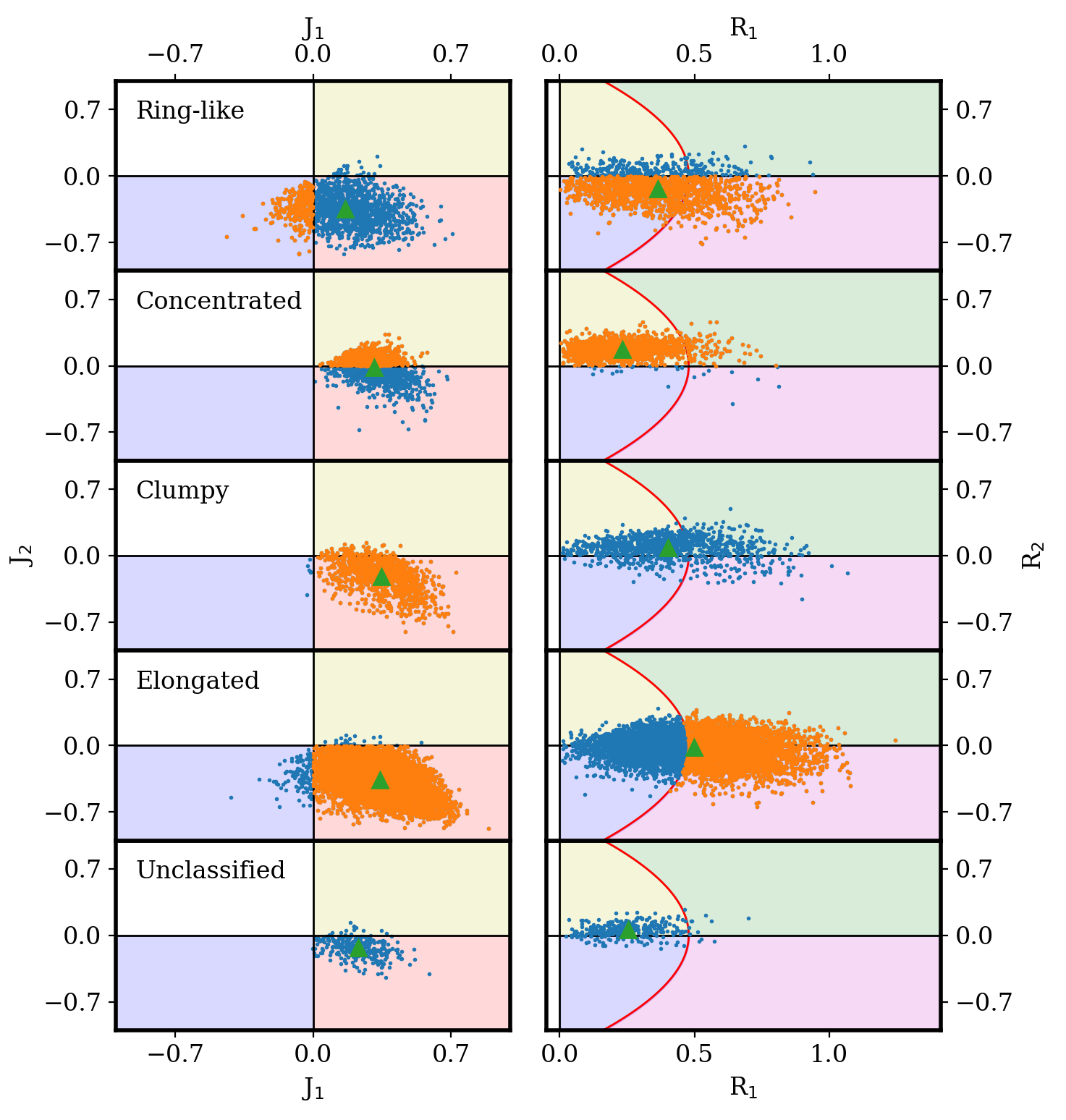}
\caption{(Left column) J-plots and (right column) RJ-plots of the 10,663 clouds identified by \citet{Ner22}, split into their 5 visually classified catagories: ring-like clouds, concentrated clouds, clumpy clouds, elongated clouds and unclassified clouds. Clouds where the automated and visual classification methods agree are shown in orange, where they disagree they are shown in blue. The green triangle shows the mean value for the set as a whole.}
\label{fig::Cat_classes}
\end{figure*}  

\begin{figure*}
\centering
\includegraphics[width=0.79\linewidth]{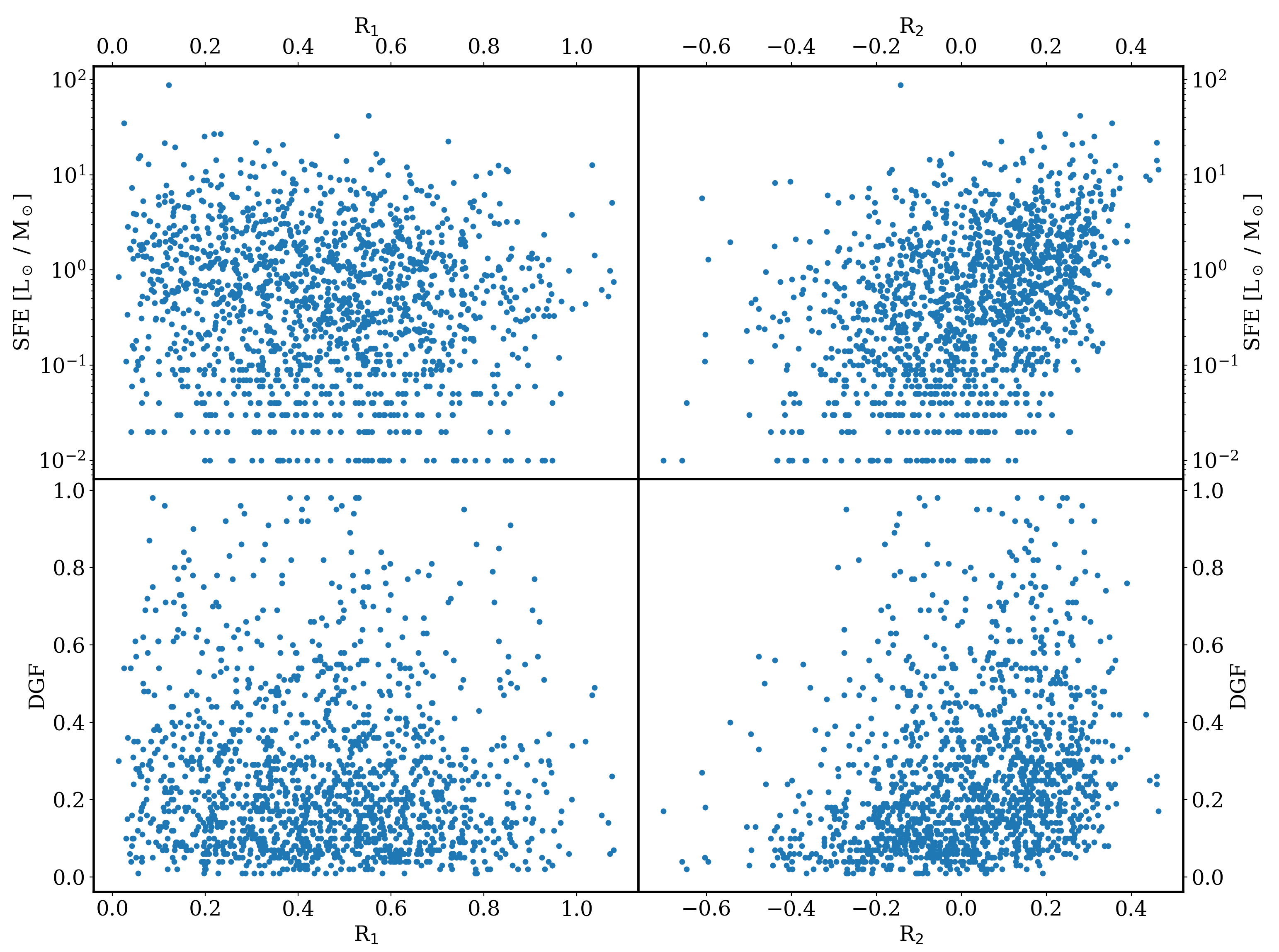}
\caption{Correlation plots between star formation efficiency (SFE, top row) and dense gas fraction (DGF, bottom row), and the RJ-values R$_1$ (left column) and R$_2$ (right column). }
\label{fig::Corr}
\end{figure*}  

\subsection{Cloud morphology - SEDIGISM}\label{SSEC:SED}%
The Structure, Excitation and Dynamics of the Inner Galactic InterStellar Medium (SEDIGISM) survey was performed using the Atacama Pathfinder Experiment (APEX) telescope from 2013-2016 \citep{Sch17,Sch21}. The survey covered the range -60$\degree$ $\leq$ $l$ $\leq$ 18$\degree$ and $|b|$ $\leq$ 0.5$\degree$, using the molecular emission lines $^{13}$CO and C$^{18}$O (J=2-1). Here we use the SEDIGISM catalogues from \citet{Dua21}, \citet{Urq21} and \citet{Ner22}, which together provide a sample of 10,663 clouds with associated properties such as dense gas fraction, as well as their J-values and visually determined morphology classification.

\subsubsection{Cloud classification and comparison to by-eye classification}\label{SSSEC:CLA}%
\citet{Ner22} classified each of the 10,663 by eye into five categories: ring-like clouds, i.e. resembling a ring, bubble or partial ring; elongated clouds, i.e. a roughly uniform but elongated structure; concentrated clouds, i.e. quasi-circular clouds with a high density region at its centre; clumpy clouds, i.e. clouds containing multiple clumps; and unclassified clouds, i.e. clouds which do not fall into any of the previous four categories. \citet{Ner22} also calculate the J-values of each cloud and find fair agreement ($78\%$) between the visual classification and J-plot classification; however, the bias for J-plots to classify structures as filaments is evident with 87$\%$ of structures classified as such by J-plots while only 57$\%$ are classified as elongated by eye. Table \ref{tab::Ner22} shows the relationship between visual and J-plot classifications. Note that \citet{Ner22} consider a visually identified clumpy cloud assigned as a J-filament or J-core to be an agreement.

\begin{table}
\centering
\begin{tabular}{lcccc}
\hline\hline
Visual  & J-plots & RJ-plots & $\overline{\mathrm{R}}_1$ & $\overline{\mathrm{R}}_2$ \\
classification& agreement & agreement & &\\ \hline
Ring-like & 17.6$\%$ & 81.7\% & 0.364 & -0.133\\
Concentrated & 50.6\% & 98.5\% & 0.233 & 0.176 \\
Clumpy & 99.4\% & - & 0.401 & 0.087 \\
Elongated & 97.7\% & 54.2\% & 0.497 & -0.023 \\
Unclassified & - & - & 0.255 & 0.058 \\\hline \hline
\end{tabular}
\centering
\caption{A table showing the agreement between visual classification and J/RJ-plots classification, as well as the average RJ-values for each classification.}
\label{tab::Ner22}
\end{table}

Figure \ref{fig::Cat_RJ} shows a J-plot considering all 10,663 clouds in the sample and the equivalent RJ-plot. One can see the wedge shape in the J-plot identified in section \ref{SSEC:BASIC} as elongation leads to structures lying along J$_2$ = -J$_1$ and with increasing elongation larger over/under central densities are needed to move away from this line. In the RJ-plot one sees R$_1$ values (elongation) up to $\sim 1$ and R$_2$ values ranging from -0.5 to 0.5. There is a slight trend of decreasing R$_2$ with increasing R$_1$, likely due to filament curvature becoming more readily apparent with increasing elongation. In total, 39.6\% of clouds are identified as elongated (R$_1$ $>$ R$_1^E$) as opposed to quasi-circular, and 54.8\% are identified as centrally over-dense (R$_2$ $>$ 0)

To compare the agreement between by-eye classification and J/RJ-plots classification, we show the J- and RJ-plots of clouds split by their by-eye classification in figure \ref{fig::Cat_classes}; where there is agreement the cloud is coloured orange and where there is disagreement it is coloured blue. The agreement between visual and RJ-plot classification is shown in a quantified manner in table \ref{tab::Ner22}. For the J-plots we follow the criteria for agreement laid out by \citet{Ner22}: ring-like clouds should be classified as J-bubbles; concentrated clouds should be classified as J-cores; clumpy clouds should be classified as J-cores or J-filaments; and elongated clouds should be classified as J-filaments. For the RJ-plots we use the following criteria for agreement: ring-like clouds should have R$_2$ $<$ 0; concentrated clouds should have R$_2$ $>$ 0; and elongated clouds must have R$_1$ $>$ R$_1^E$. For clumpy clouds there is no clear moment of inertia based definition as neither R$_1$ or R$_2$ necessarily contain information about the level of fragmentation in a structure. Rather, a separate method such as the number of leaves in a dendrogram might be sensitive to the fragmentation in a structure \citep[see for example][]{Cla18}. However, an exploration of such a method is beyond the scope of this paper and we exclude from our analysis the classification of clumpy structures. 

In figure \ref{fig::Cat_classes} and table \ref{tab::Ner22} we see a considerable improvement in the classification of ring-like clouds and concentrated clouds, going from 17.6\% and 50.6\% agreement to 81.7\% and 98.5\% agreement respectively. Elongated clouds see a decrease in agreement as RJ-plots places a more stringent requirement on this classification than J-plots. In J-plots any aspect ratio greater than 1, without considerable central over/under-density, will result in the classification of a structure as a J-filament, while RJ-plots requires A $>$ 2. The large number of visually classified elongated clouds which lie below this aspect ratio likely results from the subjective nature of by-eye classification, as well as the fact that there does not exist a clear by-eye classification category in \citet{Ner22} for a quasi-circular, roughly uniform structure. Taken together, RJ-plots provide a more robust automated method for classification than J-plots and eliminates the strong bias of J-plots to classify structures as filaments.  

In addition to looking at discrete classifications using RJ-plots, one may consider the average R$_1$ and R$_2$ of each category (also shown in table \ref{tab::Ner22}). From this, one can see that the four main cloud categories are relatively distinct from each other, even though each category shows a spread in RJ-space. Moreover, one sees that ring-like and concentrated clouds are both quasi-circular on average, and this is also true for clumpy and unclassified clouds. In general, only ring-like and concentrated clouds show clear signs of being centrally under/over-dense respectively, while elongated, clumpy and unclassified clouds have R$_2 \sim$ 0. Even though a moments of inertia based classification scheme cannot confirm the identification of clumpy clouds, one can still investigate their distribution in RJ-space. Doing so, we find that clumpy clouds, while having an average R$_2$ value close to zero, predominately have R$_2$ $>$ 0 ($\sim81\%$); suggesting that the clumps do not lie far from the cloud centre but are predominately uniformally distributed or slightly centrally concentrated. 

\begin{figure*}
\centering
\includegraphics[width=0.79\linewidth]{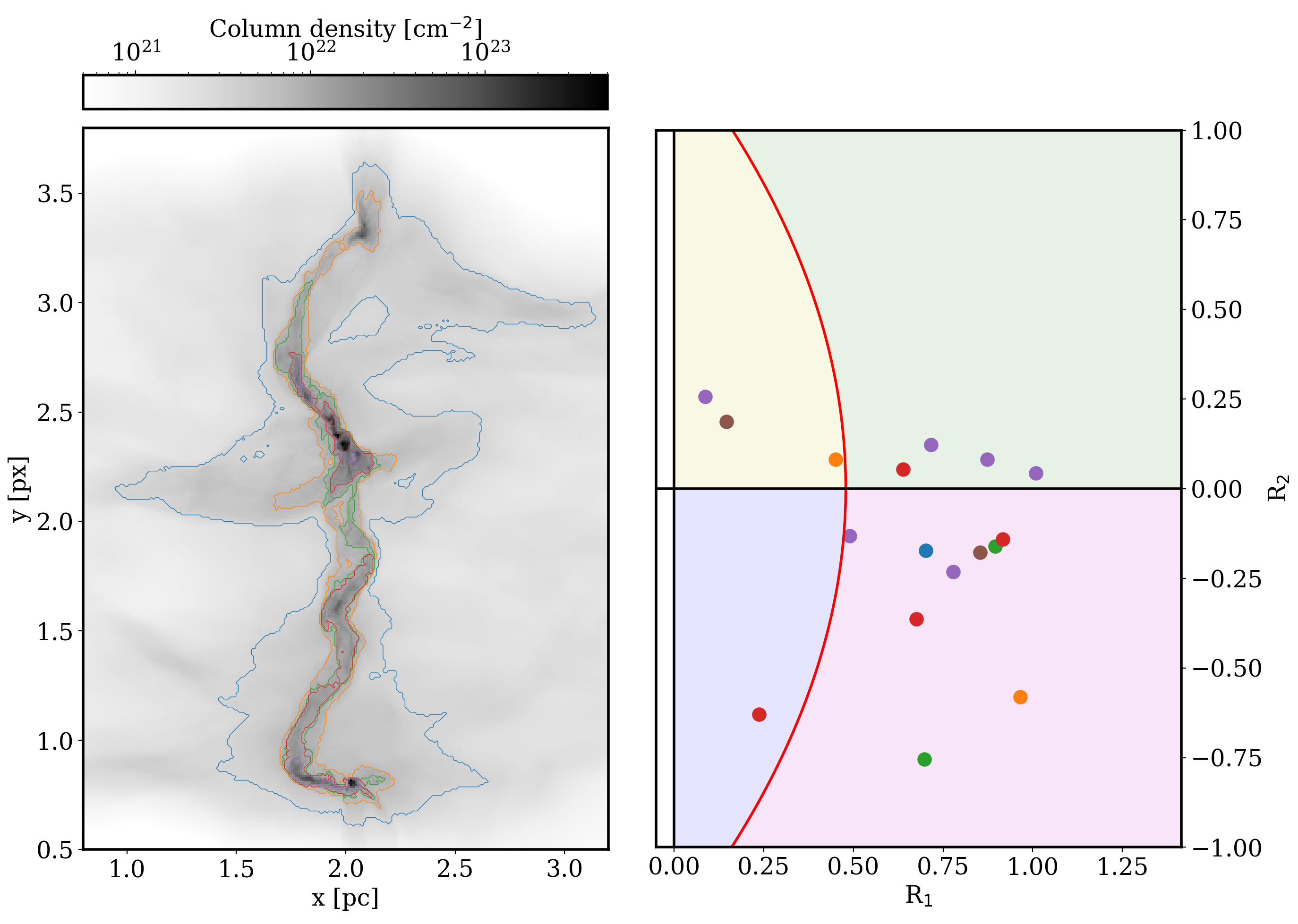}
\caption{(Left) A column density plot of a simulation from \citet{Cla20}. The contours denote the dendrogram structures identified, each colour shows the level of the structure in the dendrogram going from blue (lowest column-density), through orange, green, red and lilac, to brown (highest column-density). (Right) A RJ-plot of the dendrogram identified structures colour-coded using the same colours as the left panel.}
\label{fig::Sim1}
\end{figure*}  

\begin{figure*}
\centering
\includegraphics[width=0.79\linewidth]{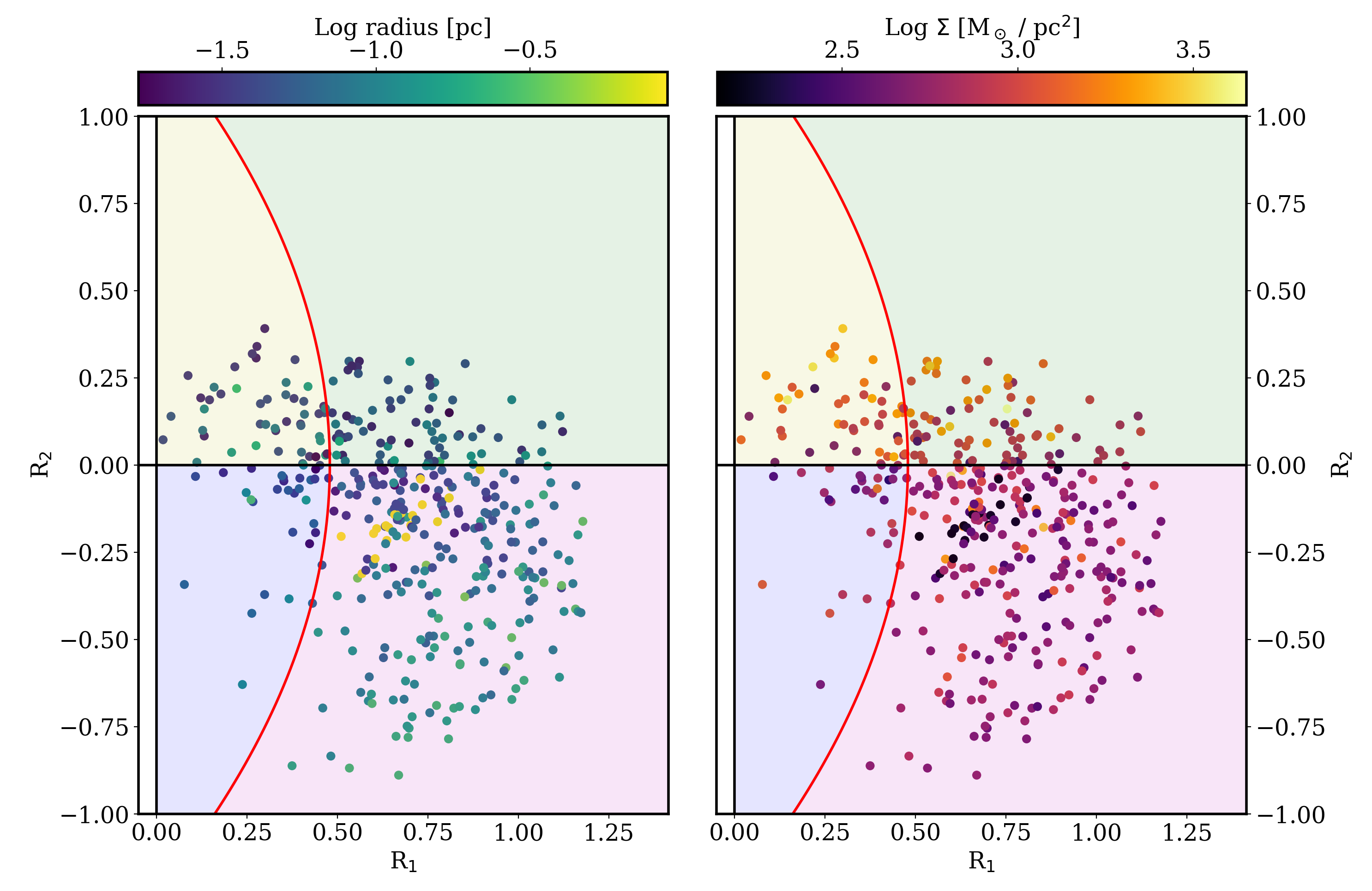}
\caption{RJ-plots of the structures identified in all 10 simulations and both viewing angles. Colour-coded are (left) the equivalent radius of the structure and (right) the average surface density of the structure.}
\label{fig::Sim_structs}
\end{figure*}  

\begin{figure*}
\centering
\includegraphics[width=0.33\linewidth]{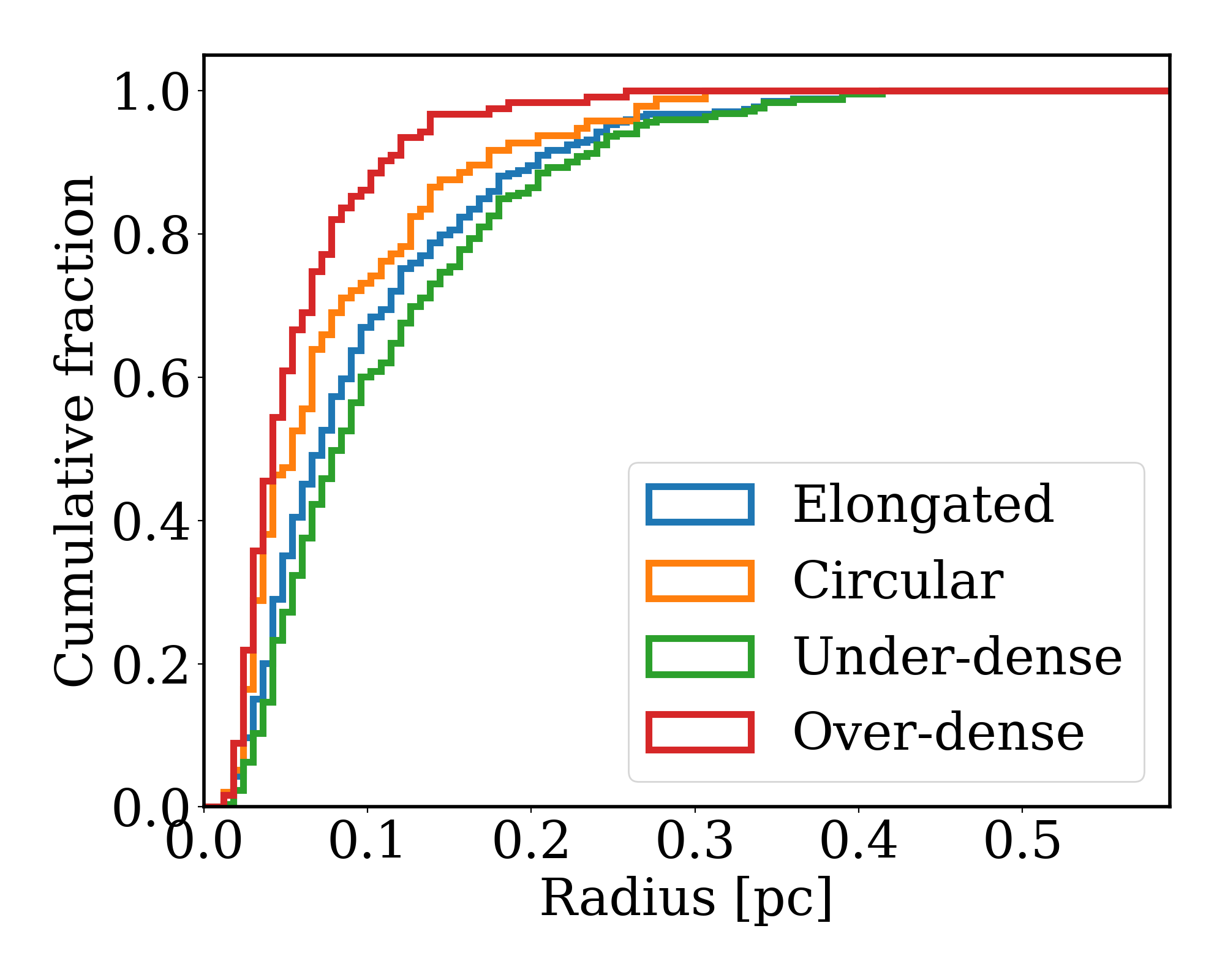}
\includegraphics[width=0.33\linewidth]{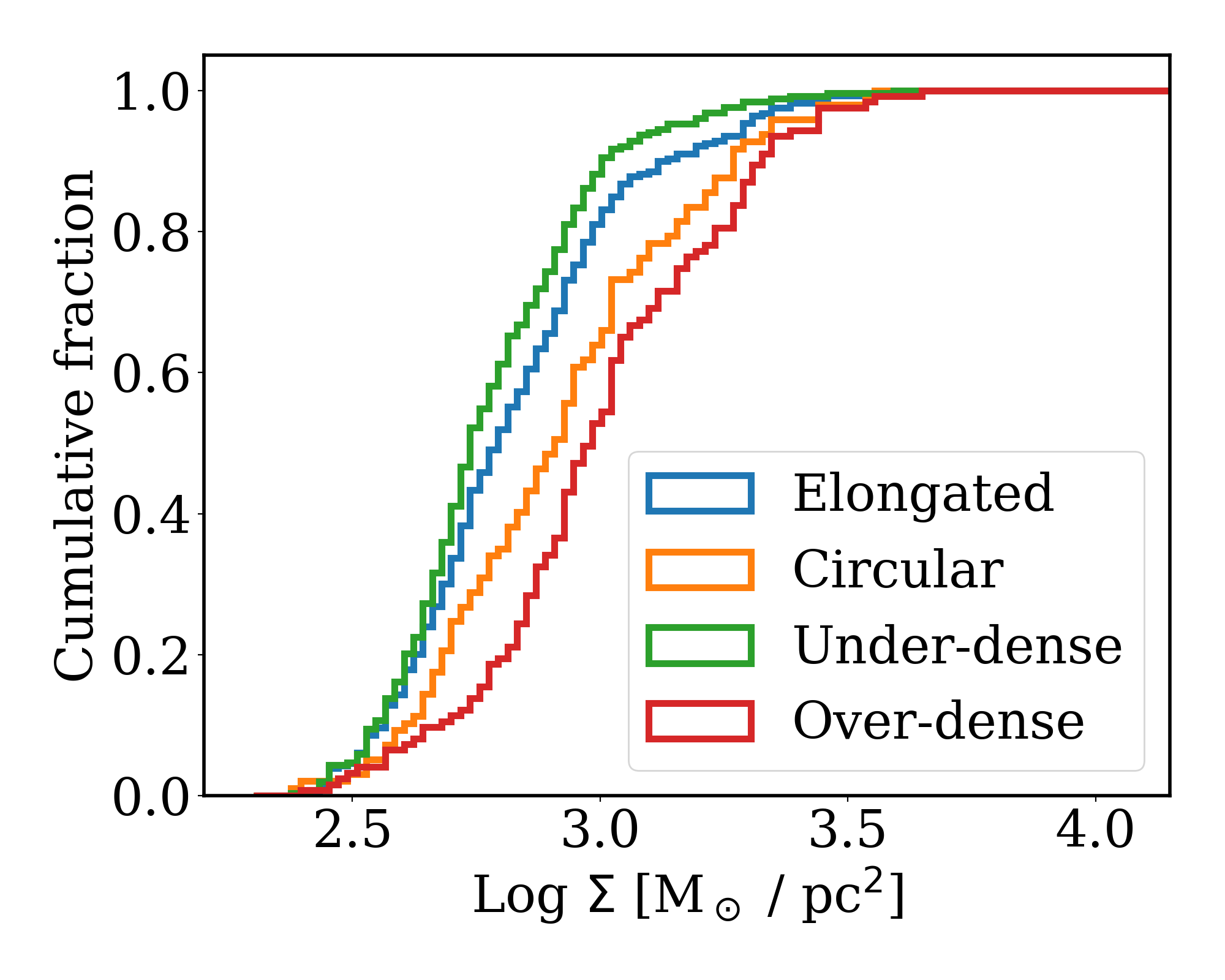}
\includegraphics[width=0.33\linewidth]{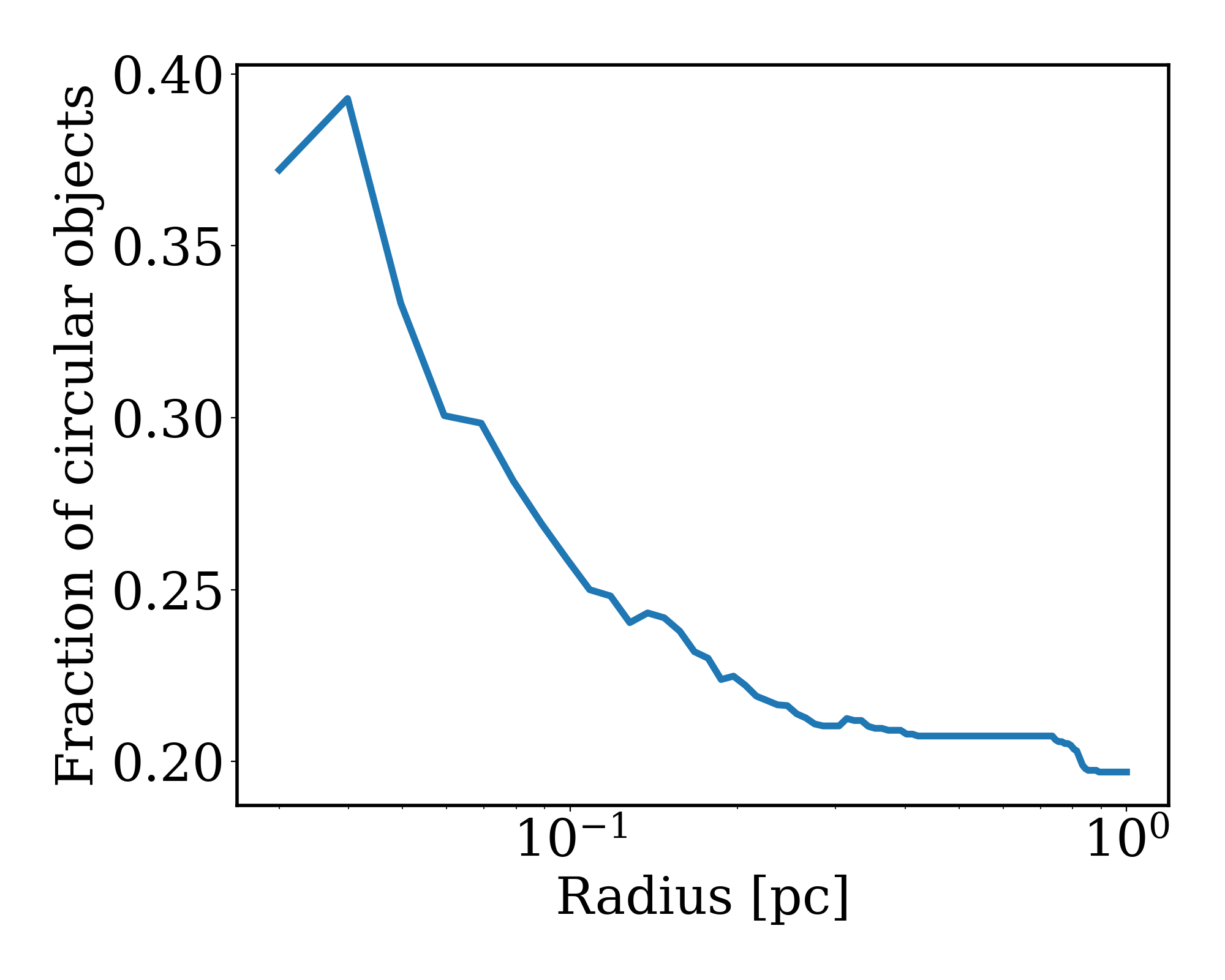}
\caption{(Left) A cumulative distribution of the equivalent radii of structures split into two sets of binary classifications: elongated vs quasi-circular, and centrally over-dense vs. under-dense. (Middle) A cumulative distribution of average surface densities with the same split as the left panel. (Right) The fraction of structures with sizes less than a given radius classified as quasi-circular.}
\label{fig::prop}
\end{figure*}  

\subsubsection{Cloud morphology and physical quantities}\label{SSSEC:PQ}%
\citet{Ner22} use null hypothesis tests to look for statistically significant differences in the dense gas fraction and star formation efficiency of clouds between the morphology classes. This however relies on the rather arbitrary definitions used to classify clouds into discrete categories when in reality morphology is a continuum, as seen in the spread in RJ-space within each visual classification category. Using the fact that RJ-values approximately disentangle cloud elongation from cloud central concentration, one may use simple correlations between R$_1$ and R$_2$, and cloud properties to investigate the link between these morphological traits and those properties. We show such correlations with dense gas fraction (DGF) and star formation efficiency (SFE) from the SEDIGISM sample in figure \ref{fig::Corr}; note that only 1,518 clouds from the total 10,663 have non-zero SFE and DGF and a DGF $<$ 1.  

Looking at figure \ref{fig::Corr}, one sees little correlation between either SFE or DGF with R$_1$, showing that a cloud's elongation appears to have no effect on the amount of dense gas or its star formation potential. This is confirmed by using the Kendall $\tau$ correlation test \citep{Ken38} which finds only minor, though statistically significant, negative correlations, $\tau\sim-0.05$, $p\sim0.001$. Considering correlations with R$_2$, both SFE and DGF present clear positive correlations, $\tau\sim0.25$, $p\ll 10^{-5}$. Thus, the clouds which are centrally concentrated, but not necessarily with only a single clump, are characterised by higher star formation efficiencies and dense gas fractions than those which are under-dense at their centre. This agrees with the result using null hypothesis testing from \citet{Ner22} which found that ring-like clouds have the lowest average SFE and DGF compared to concentrated and clumpy clouds which posses the highest. However, \citet{Ner22} do find elongated clouds to have lower than average SFE and DGF; our results show that this is likely due to elongated clouds being centrally under-dense in general ($\overline{\mathrm{R}}_2$=-0.023), and is actually unrelated to their elongation. 

\subsection{Multi-scale morphology - simulations}\label{SSEC:SIM}%
To highlight a multi-scale approach to the use of RJ-plots we use simulation data from \citet{Cla20} to investigate the morphology of sub-structures within a non-equilibrium, accreting filament. These simulations were performed using the moving-mesh code \textsc{Arepo} \citep{Spr10}. The code solves the three-dimensional hydrodynamic equations including self-gravity with time-dependent coupled chemistry and thermodynamics. The initial set-up is a cylindrical, turbulent colliding flow; the set-up quickly forms a dense filament at the $z$-axis. In total 10 simulations were performed with identical properties, only the random seed used to generate the turbulent velocity field being changed. Accretion from the colliding flows drives internal turbulence and the filaments fragment to form numerous cores as well as filamentary sub-structures \citep[see][for the formation process and properties of these sub-structures]{Cla17,Cla18,Cla20}. The simulations are stopped when $\sim$5 - 10$\%$ of the mass of the filament is contained in sink particles and at this point the simulations are analysed. This occurs about 0.5 - 0.6 Myr after the start of the simulation and the filaments have reached a mass between $\sim$150 M$\Su$ and $\sim$250 M$\Su$. Prior to analysis, the \textsc{Arepo} Voronoi-mesh is mapped to a fixed Cartesian grid with resolution of 0.01 pc. For each simulation we use two viewing angles, those aligned with the $x$- and $y$-axes, to produce a total of 20 column density plots. An example column density image is shown in figure \ref{fig::Sim1}. For full details on the simulations see \citet{Cla18} and \citet{Cla20}.

To identify structures on multiple scales we use a dendrogram approach, specifically the \textsc{Python} package \textsc{astrodendro}\footnote{http://www.dendrograms.org/}. The dendrograms are constructed on the logarithmic column density as the images cover a large dynamic range. The parameters used are: \textsc{min$\_$value} = 21.5, \textsc{min$\_$delta} = 0.3, \textsc{min$\_$npix} = 9. This value of \textsc{min$\_$value} is adopted as it is sufficiently high as to only include the inner filament region of the map but sufficiently low as to ensure that there is only one single dendrogram trunk, i.e. a single structure on the largest scale. The value of \textsc{min$\_$delta} = 0.3 corresponds to roughly a doubling in linear space; increasing its value will decrease the number of structures found between the dendrogram trunk and its leaves, while decreasing will increase this number. We consider 0.3 to be a good compromise value and one which does not greatly affect the results. Setting \textsc{min$\_$npix} = 9 places a minimum diameter of a structure of $\sim$0.03 pc and ensures that any structure is at least marginally resolved. The left panel of figure \ref{fig::Sim1} shows the dendrogram structures identified in this manner for simulation S1 as contours, the colour of the contour denoting a structure's level in the dendrogram. 

We calculate the RJ-values of each of the structures; this is done using the linear column density image of each structure. The RJ-plot generated from considering the structures in simulation S1 can be seen in the right panel of figure \ref{fig::Sim1}. There one can see that the majority of the structures are elongated, with few lying to the left of the red boundary line. Further, higher level dendrogram structures (i.e. smaller and more deeply embedded) are more likely to be concentrated, while lower level structures appear as curved elongated structures (R$_1$ $>$ R$_1^E$, R$_2$ $<$ 0), showing a distinction between the larger-scale filaments and the smaller-scale cores. 

Considering all 20 column density maps results in 396 total structures with masses ranging from 0.36 to 354.3 M$\Su$ and equivalent radii from 0.02 - 0.88 pc. A structure's equivalent radius is defined as $r = \sqrt{a/\pi}$, where $a$ is the structure's area; this is only approximate due to the non-circular nature of the structures but may be used to estimate the size of a structure. Figure \ref{fig::Sim_structs} shows the RJ-plots containing all 396 structures, colour-coded by the structure size and average surface density. One can see that the majority of structures are elongated ($\sim$80\%) as well as the fact that there exists two trends in RJ-space. The first set of structures extends vertically between 0.5 $\lesssim$ R$_1$ $\lesssim$ 1.1 and have relatively large radii (r $\sim$ 0.4 pc) and moderate surface densities ($\Sigma \sim 500$ M/pc$^2$). These structures lie in the same region as the sectors of an annulus, as considered in figure \ref{fig::rings}, and are regions of the filament which are highly curved. As an example see the southern section of the filament in figure \ref{fig::Sim1}. The second set of structures lie close to R$_2$ $\sim$ 0 but show a clear anti-correlation between R$_1$ and R$_2$, i.e. less elongated structures are more centrally concentrated. These structures are typically small (r $\lesssim$ 0.1 pc), have high surface densities ($\Sigma \gtrsim 1000$ M/pc$^2$), and are associated with the small core structures in the filament.

These trends across scales can be seen in figure \ref{fig::prop} which shows the cumulative distributions of sizes and average surface densities of structures split into two sets of binary classifications: elongated vs quasi-circular, and centrally over-dense vs. under-dense. Quasi-circular structures are both smaller and denser than elongated structures, and the same is true of centrally over-dense structures compared to centrally under-dense structures. In the right panel of figure \ref{fig::prop} is shown the fraction of structures below a given equivalent radius which are quasi-circular compared to elongated. As one moves to smaller scales, quasi-circular structures become more common (reaching $\sim40\%$ at 0.03 pc)\footnote{At this small scale, structures contain approximately 30 pixels. Such a reduced number of pixels may affect measurements of the central concentration due to the central area being poorly resolved; however, it ought to have minimal effect on the elongation as a range of aspect ratios can still be achieved (i.e. 1-30).} but never become the dominant structure morphology. Thus, considering fragmenting filaments, structures predominately remain filamentary across all scales and core-like structures inherit their elongation from the parental filament.

\section{Conclusions}\label{SEC:CON}%
Due to the wealth of data from observations and simulations an automated method for the quantification and classification of structure morphology is sorely needed. In this paper we presented such an automated method called RJ-plots which builds upon the work of \citet{Jaf18}. Testing this method against readily parametrisable shapes, we show that RJ-plots approximately separates a structure's elongation from its central over/under-density. This allows discrete classification of structures as quasi-circular vs elongated (R$_1$ $<$ R$_1^E$ vs. R$_1$ $>$ R$_1^E$) and as centrally over-dense vs under-dense (R$_2$ $>$ 0 vs R$_2$ $<$ 0). Centrally over-dense structures are defined as structures where a greater share of their mass/intensity/weight is close to their centre, i.e. a concentrated clump/core or hub-filament system, and centrally under-dense structure are defined as the inverse, i.e. a bubble shell, either full or partial, or if elongated, a curved filament. Beyond discrete classification, RJ-plots also allows meaningful correlations to be sought between RJ-values (R$_1$,R$_2$) and structure properties to investigate how morphology affects those properties.

When using the SEDIGISM cloud morphology catalogue of \citet{Ner22}, we find a significant improvement in the agreement between the visual classification of structures and that provided by RJ-plots when compared to J-plots. This is most markedly seen in the visually classified ring-like and concentrated clouds, going from 17.6\% and 50.6\% agreement with J-plots to 81.7\% and 98.5\% agreement with RJ-plots respectively. Moreover, by studying the relationship between a cloud's star formation efficiency and dense gas fraction with RJ-values, we find a strong positive correlation between a cloud's central concentration and these quantities, as well as a lack of correlation with a cloud's elongation.

We use column density maps from the simulations of \citet{Cla20} to investigate the morphology of sub-structures across multiple scales within a non-equilibrium, accreting filament. There we find that large sections of the filaments are highly curved (R$_2\sim-0.5$), while on smaller scales, structures become increasingly spherical and centrally over-dense. We show that while quasi-circular structures become more common at smaller scales, they are never the dominant structure morphology. Rather, a filament's elongation is inherited by its sub-structures down to the core scale (r $\sim0.03$ pc).

We provide an open-source code for calculating RJ-values and automatically classifying structures at https://github.com/SeamusClarke/RJplots

\section{Acknowledgments}\label{SEC:ACK}%
We thank the referee for their helpful comments which helped clarify parts of the manuscript. SDC thanks Ya-Wen Tang and Patrick Koch for constructive discussions which helped with the focus of this work. SDC is supported by the Ministry of Science and Technology (MoST) in Taiwan through grant MoST 108-2112-M-001-004-MY2. APW gratefully acknowledges the support of a consolidated grant (ST/K00926/1) from the UK Science and Technology Facilities Council. SDC would like to thank Kartik Rajan Neralwar for providing access to their morphology catalogue shortly prior to it becoming public. This publication uses data acquired with the Atacama Pathfinder Experiment (APEX) telescope under programmes 092.F-9315 and 193.C-0584. APEX is a collaboration among the Max-Planck-Institut fur Radioastronomie, the European Southern Observatory, and the Onsala Space Observatory. The processed data products are available from the SEDIGISM survey database located at https://sedigism.mpifr-bonn.mpg.de/index.html, which was constructed by James Urquhart and hosted by the Max Planck Institute for Radio Astronomy. SDC would also like to acknowledge the people developing and maintaining the open source packages which were used in this work: \textsc{Matplotlib} \citep{matplotlib}, \textsc{NumPy} \citep{numpy} and \textsc{SciPy} \citep{scipy}.

\section{Data availability}\label{SEC:DATA}%
The SEDIGISM data used in this article is publicly available and can be found at https://sedigism.mpifr-bonn.mpg.de/index.html. The \citet{Cla20} simulation data used in this article will be shared on reasonable request to the corresponding author. The code for determining RJ-values, automated structure classification and plotting is public and can be found at https://github.com/SeamusClarke/RJplots

\bibliographystyle{mn2e}
\bibliography{ref} 

\begin{thebibliography}{35}
\expandafter\ifx\csname natexlab\endcsname\relax\def\natexlab#1{#1}\fi

\bibitem[{{Andr{\'e}} {et~al}\mbox{.}(2010){Andr{\'e}}, {Men'shchikov},
  {Bontemps}, {K{\"o}nyves}, {Motte}, {Schneider}, {Didelon}, {Minier},
  {Saraceno}, {Ward-Thompson}, {di Francesco}, {White}, {Molinari}, {Testi},
  {Abergel}, {Griffin}, {Henning}, {Royer}, {Mer{\'{\i}}n}, {Vavrek}, {Attard},
  {Arzoumanian}, {Wilson}, {Ade}, {Aussel}, {Baluteau}, {Benedettini},
  {Bernard}, {Blommaert}, {Cambr{\'e}sy}, {Cox}, {di Giorgio}, {Hargrave},
  {Hennemann}, {Huang}, {Kirk}, {Krause}, {Launhardt}, {Leeks}, {Le Pennec},
  {Li}, {Martin}, {Maury}, {Olofsson}, {Omont}, {Peretto}, {Pezzuto}, {Prusti},
  {Roussel}, {Russeil}, {Sauvage}, {Sibthorpe}, {Sicilia-Aguilar}, {Spinoglio},
  {Waelkens}, {Woodcraft}, \& {Zavagno}}]{And10}
{Andr{\'e}} P. {et~al.}, 2010, \aap, 518, L102

\bibitem[{{Arzoumanian} {et~al}\mbox{.}(2019){Arzoumanian}, {Andr{\'e}},
  {K{\"o}nyves}, {Palmeirim}, {Roy}, {Schneider}, {Benedettini}, {Didelon}, {Di
  Francesco}, {Kirk}, \& {Ladjelate}}]{Arz19}
{Arzoumanian} D. {et~al.}, 2019, \aap, 621, A42

\bibitem[{{Berry}(2015)}]{Ber15}
{Berry} D.~S., 2015, Astronomy and Computing, 10, 22

\bibitem[{{Churchwell} {et~al}\mbox{.}(2006){Churchwell}, {Povich}, {Allen},
  {Taylor}, {Meade}, {Babler}, {Indebetouw}, {Watson}, {Whitney}, {Wolfire},
  {Bania}, {Benjamin}, {Clemens}, {Cohen}, {Cyganowski}, {Jackson},
  {Kobulnicky}, {Mathis}, {Mercer}, {Stolovy}, {Uzpen}, {Watson}, \&
  {Wolff}}]{Chu06}
{Churchwell} E. {et~al.}, 2006, \apj, 649, 759

\bibitem[{{Clarke} {et~al}\mbox{.}(2017){Clarke}, {Whitworth}, {Duarte-Cabral},
  \& {Hubber}}]{Cla17}
{Clarke} S.~D., {Whitworth} A.~P., {Duarte-Cabral} A., {Hubber} D.~A., 2017,
  \mnras, 468, 2489

\bibitem[{{Clarke} {et~al}\mbox{.}(2018){Clarke}, {Whitworth}, {Spowage},
  {Duarte-Cabral}, {Suri}, {Jaffa}, {Walch}, \& {Clark}}]{Cla18}
{Clarke} S.~D., {Whitworth} A.~P., {Spowage} R.~L., {Duarte-Cabral} A., {Suri}
  S.~T., {Jaffa} S.~E., {Walch} S., {Clark} P.~C., 2018, \mnras

\bibitem[{{Clarke}, {Williams} \& {Walch}(2020){Clarke}, {Williams}, \&
  {Walch}}]{Cla20}
{Clarke} S.~D., {Williams} G.~M., {Walch} S., 2020, \mnras, 497, 4390

\bibitem[{{Colombo} {et~al}\mbox{.}(2021){Colombo}, {K{\"o}nig}, {Urquhart},
  {Wyrowski}, {Mattern}, {Menten}, {Lee}, {Brand}, {Wienen}, {Mazumdar},
  {Schuller}, \& {Leurini}}]{Col21}
{Colombo} D. {et~al.}, 2021, \aap, 655, L2

\bibitem[{{Colombo} {et~al}\mbox{.}(2015){Colombo}, {Rosolowsky}, {Ginsburg},
  {Duarte-Cabral}, \& {Hughes}}]{Col15}
{Colombo} D., {Rosolowsky} E., {Ginsburg} A., {Duarte-Cabral} A., {Hughes} A.,
  2015, \mnras, 454, 2067

\bibitem[{{Daigle}, {Joncas} \& {Parizeau}(2007){Daigle}, {Joncas}, \&
  {Parizeau}}]{Dai07}
{Daigle} A., {Joncas} G., {Parizeau} M., 2007, \apj, 661, 285

\bibitem[{{Duarte-Cabral} {et~al}\mbox{.}(2021){Duarte-Cabral}, {Colombo},
  {Urquhart}, {Ginsburg}, {Russeil}, {Schuller}, {Anderson}, {Barnes},
  {Beltr{\'a}n}, {Beuther}, {Bontemps}, {Bronfman}, {Csengeri}, {Dobbs},
  {Eden}, {Giannetti}, {Kauffmann}, {Mattern}, {Medina}, {Menten}, {Lee},
  {Pettitt}, {Riener}, {Rigby}, {Traficante}, {Veena}, {Wienen}, {Wyrowski},
  {Agurto}, {Azagra}, {Cesaroni}, {Finger}, {Gonzalez}, {Henning}, {Hernandez},
  {Kainulainen}, {Leurini}, {Lopez}, {Mac-Auliffe}, {Mazumdar}, {Molinari},
  {Motte}, {Muller}, {Nguyen-Luong}, {Parra}, {Perez-Beaupuits},
  {Montenegro-Montes}, {Moore}, {Ragan}, {S{\'a}nchez-Monge}, {Sanna},
  {Schilke}, {Schisano}, {Schneider}, {Suri}, {Testi}, {Torstensson},
  {Venegas}, {Wang}, \& {Zavagno}}]{Dua21}
{Duarte-Cabral} A. {et~al.}, 2021, \mnras, 500, 3027

\bibitem[{{Ehlerov{\'a}} \& {Palou{\v{s}}}(2013)}]{Ehl13}
{Ehlerov{\'a}} S., {Palou{\v{s}}} J., 2013, \aap, 550, A23

\bibitem[{Harris {et~al}\mbox{.}(2020)Harris, Millman, van~der Walt, Gommers,
  Virtanen, Cournapeau, Wieser, Taylor, Berg, Smith, Kern, Picus, Hoyer, van
  Kerkwijk, Brett, Haldane, del R{\'{i}}o, Wiebe, Peterson,
  G{\'{e}}rard-Marchant, Sheppard, Reddy, Weckesser, Abbasi, Gohlke, \&
  Oliphant}]{numpy}
Harris C.~R. {et~al.}, 2020, Nature, 585, 357

\bibitem[{Hunter(2007)}]{matplotlib}
Hunter J.~D., 2007, Computing in Science and Engineering, 9, 90

\bibitem[{{Jaffa} {et~al}\mbox{.}(2018){Jaffa}, {Whitworth}, {Clarke}, \&
  {Howard}}]{Jaf18}
{Jaffa} S.~E., {Whitworth} A.~P., {Clarke} S.~D., {Howard} A.~D.~P., 2018,
  \mnras

\bibitem[{Kendall(1938)}]{Ken38}
Kendall M.~G., 1938, Biometrika, 30, 81

\bibitem[{{Li} {et~al}\mbox{.}(2016){Li}, {Urquhart}, {Leurini}, {Csengeri},
  {Wyrowski}, {Menten}, \& {Schuller}}]{Li16}
{Li} G.-X., {Urquhart} J.~S., {Leurini} S., {Csengeri} T., {Wyrowski} F.,
  {Menten} K.~M., {Schuller} F., 2016, \aap, 591, A5

\bibitem[{{McClure-Griffiths} {et~al}\mbox{.}(2002){McClure-Griffiths},
  {Dickey}, {Gaensler}, \& {Green}}]{McC02}
{McClure-Griffiths} N.~M., {Dickey} J.~M., {Gaensler} B.~M., {Green} A.~J.,
  2002, \apj, 578, 176

\bibitem[{{Men'shchikov}(2021)}]{Men21}
{Men'shchikov} A., 2021, \aap, 649, A89

\bibitem[{{Men'shchikov} {et~al}\mbox{.}(2012){Men'shchikov}, {Andr{\'e}},
  {Didelon}, {Motte}, {Hennemann}, \& {Schneider}}]{Men12}
{Men'shchikov} A., {Andr{\'e}} P., {Didelon} P., {Motte} F., {Hennemann} M.,
  {Schneider} N., 2012, \aap, 542, A81

\bibitem[{{Molinari} {et~al}\mbox{.}(2010){Molinari}, {Swinyard}, {Bally},
  {Barlow}, {Bernard}, {Martin}, {Moore}, {Noriega-Crespo}, {Plume}, {Testi},
  {Zavagno}, {Abergel}, {Ali}, {Anderson}, {Andr{\'e}}, {Baluteau},
  {Battersby}, {Beltr{\'a}n}, {Benedettini}, {Billot}, {Blommaert}, {Bontemps},
  {Boulanger}, {Brand}, {Brunt}, {Burton}, {Calzoletti}, {Carey}, {Caselli},
  {Cesaroni}, {Cernicharo}, {Chakrabarti}, {Chrysostomou}, {Cohen},
  {Compiegne}, {de Bernardis}, {de Gasperis}, {di Giorgio}, {Elia}, {Faustini},
  {Flagey}, {Fukui}, {Fuller}, {Ganga}, {Garcia-Lario}, {Glenn}, {Goldsmith},
  {Griffin}, {Hoare}, {Huang}, {Ikhenaode}, {Joblin}, {Joncas}, {Juvela},
  {Kirk}, {Lagache}, {Li}, {Lim}, {Lord}, {Marengo}, {Marshall}, {Masi},
  {Massi}, {Matsuura}, {Minier}, {Miville-Desch{\^e}nes}, {Montier}, {Morgan},
  {Motte}, {Mottram}, {M{\"u}ller}, {Natoli}, {Neves}, {Olmi}, {Paladini},
  {Paradis}, {Parsons}, {Peretto}, {Pestalozzi}, {Pezzuto}, {Piacentini},
  {Piazzo}, {Polychroni}, {Pomar{\`e}s}, {Popescu}, {Reach}, {Ristorcelli},
  {Robitaille}, {Robitaille}, {Rod{\'o}n}, {Roy}, {Royer}, {Russeil},
  {Saraceno}, {Sauvage}, {Schilke}, {Schisano}, {Schneider}, {Schuller},
  {Schulz}, {Sibthorpe}, {Smith}, {Smith}, {Spinoglio}, {Stamatellos},
  {Strafella}, {Stringfellow}, {Sturm}, {Taylor}, {Thompson}, {Traficante},
  {Tuffs}, {Umana}, {Valenziano}, {Vavrek}, {Veneziani}, {Viti}, {Waelkens},
  {Ward-Thompson}, {White}, {Wilcock}, {Wyrowski}, {Yorke}, \& {Zhang}}]{Mol10}
{Molinari} S. {et~al.}, 2010, \aap, 518, L100

\bibitem[{{Neralwar} {et~al}\mbox{.}(2022){Neralwar}, {Colombo},
  {Duarte-Cabral}, {Urquhart}, {Mattern}, {Wyrowski}, {Menten}, {Barnes},
  {Sanchez-Monge}, {Beuther}, {Rigby}, {Mazumdar}, {Eden}, {Csengeri}, {Dobbs},
  {Veena}, {Neupane}, {Henning}, {Schuller}, {Leurini}, {Wienen}, {Yang},
  {Ragan}, {Medina}, \& {Nguyen-Luong}}]{Ner22}
{Neralwar} K.~R. {et~al.}, 2022, arXiv e-prints, arXiv:2203.02504

\bibitem[{{Petkova} {et~al}\mbox{.}(2021){Petkova}, {Kruijssen}, {Kluge},
  {Glover}, {Walker}, {Longmore}, {Henshaw}, {Reissl}, \& {Dale}}]{Pet21}
{Petkova} M.~A. {et~al.}, 2021, arXiv e-prints, arXiv:2104.09558

\bibitem[{{Pineda} {et~al}\mbox{.}(2022){Pineda}, {Arzoumanian}, {Andr{\'e}},
  {Friesen}, {Zavagno}, {Clarke}, {Inoue}, {Chen}, {Lee}, {Soler}, \&
  {Kuffmeier}}]{Pin22}
{Pineda} J.~E. {et~al.}, 2022, arXiv e-prints, arXiv:2205.03935

\bibitem[{{Rigby} {et~al}\mbox{.}(2016){Rigby}, {Moore}, {Plume}, {Eden},
  {Urquhart}, {Thompson}, {Mottram}, {Brunt}, {Butner}, {Dempsey}, {Gibson},
  {Hatchell}, {Jenness}, {Kuno}, {Longmore}, {Morgan}, {Polychroni}, {Thomas},
  {White}, \& {Zhu}}]{Rig16}
{Rigby} A.~J. {et~al.}, 2016, \mnras, 456, 2885

\bibitem[{{Rosolowsky} {et~al}\mbox{.}(2008){Rosolowsky}, {Pineda},
  {Kauffmann}, \& {Goodman}}]{Ros08}
{Rosolowsky} E.~W., {Pineda} J.~E., {Kauffmann} J., {Goodman} A.~A., 2008,
  \apj, 679, 1338

\bibitem[{{Schisano} {et~al}\mbox{.}(2020){Schisano}, {Molinari}, {Elia},
  {Benedettini}, {Olmi}, {Pezzuto}, {Traficante}, {Brescia}, {Cavuoti}, {di
  Giorgio}, {Liu}, {Moore}, {Noriega-Crespo}, {Riccio}, {Baldeschi},
  {Becciani}, {Peretto}, {Merello}, {Vitello}, {Zavagno}, {Beltr{\'a}n},
  {Cambr{\'e}sy}, {Eden}, {Li Causi}, {Molinaro}, {Palmeirim}, {Sciacca},
  {Testi}, {Umana}, \& {Whitworth}}]{Sch20}
{Schisano} E. {et~al.}, 2020, \mnras, 492, 5420

\bibitem[{{Schisano} {et~al}\mbox{.}(2014){Schisano}, {Rygl}, {Molinari},
  {Busquet}, {Elia}, {Pestalozzi}, {Polychroni}, {Billot}, {Carey}, {Paladini},
  {Noriega-Crespo}, {Moore}, {Plume}, {Glover}, \&
  {V{\'a}zquez-Semadeni}}]{Sch14}
{Schisano} E. {et~al.}, 2014, \apj, 791, 27

\bibitem[{{Schuller} {et~al}\mbox{.}(2017){Schuller}, {Csengeri}, {Urquhart},
  {Duarte-Cabral}, {Barnes}, {Giannetti}, {Hernandez}, {Leurini}, {Mattern},
  {Medina}, {Agurto}, {Azagra}, {Anderson}, {Beltr{\'a}n}, {Beuther},
  {Bontemps}, {Bronfman}, {Dobbs}, {Dumke}, {Finger}, {Ginsburg}, {Gonzalez},
  {Henning}, {Kauffmann}, {Mac-Auliffe}, {Menten}, {Montenegro-Montes},
  {Moore}, {Muller}, {Parra}, {Perez-Beaupuits}, {Pettitt}, {Russeil},
  {S{\'a}nchez-Monge}, {Schilke}, {Schisano}, {Suri}, {Testi}, {Torstensson},
  {Venegas}, {Wang}, {Wienen}, {Wyrowski}, \& {Zavagno}}]{Sch17}
{Schuller} F. {et~al.}, 2017, \aap, 601, A124

\bibitem[{{Schuller} {et~al}\mbox{.}(2021){Schuller}, {Urquhart}, {Csengeri},
  {Colombo}, {Duarte-Cabral}, {Mattern}, {Ginsburg}, {Pettitt}, {Wyrowski},
  {Anderson}, {Azagra}, {Barnes}, {Beltran}, {Beuther}, {Billington},
  {Bronfman}, {Cesaroni}, {Dobbs}, {Eden}, {Lee}, {Medina}, {Menten}, {Moore},
  {Montenegro-Montes}, {Ragan}, {Rigby}, {Riener}, {Russeil}, {Schisano},
  {Sanchez-Monge}, {Traficante}, {Zavagno}, {Agurto}, {Bontemps}, {Finger},
  {Giannetti}, {Gonzalez}, {Hernandez}, {Henning}, {Kainulainen}, {Kauffmann},
  {Leurini}, {Lopez}, {Mac-Auliffe}, {Mazumdar}, {Molinari}, {Motte}, {Muller},
  {Nguyen-Luong}, {Parra}, {Perez-Beaupuits}, {Schilke}, {Schneider}, {Suri},
  {Testi}, {Torstensson}, {Veena}, {Venegas}, {Wang}, \& {Wienen}}]{Sch21}
{Schuller} F. {et~al.}, 2021, \mnras, 500, 3064

\bibitem[{{Sousbie}(2011)}]{Sou11}
{Sousbie} T., 2011, \mnras, 414, 350

\bibitem[{{Springel}(2010)}]{Spr10}
{Springel} V., 2010, \mnras, 401, 791

\bibitem[{{Stutzki} \& {Guesten}(1990)}]{Stu90}
{Stutzki} J., {Guesten} R., 1990, \apj, 356, 513

\bibitem[{{Urquhart} {et~al}\mbox{.}(2021){Urquhart}, {Figura}, {Cross},
  {Wells}, {Moore}, {Eden}, {Ragan}, {Pettitt}, {Duarte-Cabral}, {Colombo},
  {Schuller}, {Csengeri}, {Mattern}, {Beuther}, {Menten}, {Wyrowski},
  {Anderson}, {Barnes}, {Beltr{\'a}n}, {Billington}, {Bronfman}, {Giannetti},
  {Kainulainen}, {Kauffmann}, {Lee}, {Leurini}, {Medina}, {Montenegro-Montes},
  {Riener}, {Rigby}, {S{\'a}nchez-Monge}, {Schilke}, {Schisano}, {Traficante},
  \& {Wienen}}]{Urq21}
{Urquhart} J.~S. {et~al.}, 2021, \mnras, 500, 3050

\bibitem[{Virtanen {et~al}\mbox{.}(2020)Virtanen, Gommers, Oliphant, Haberland,
  Reddy, Cournapeau, Burovski, Peterson, Weckesser, Bright, {van der Walt},
  Brett, Wilson, Millman, Mayorov, Nelson, Jones, Kern, Larson, Carey, Polat,
  Feng, Moore, {VanderPlas}, Laxalde, Perktold, Cimrman, Henriksen, Quintero,
  Harris, Archibald, Ribeiro, Pedregosa, {van Mulbregt}, \& {SciPy 1.0
  Contributors}}]{scipy}
Virtanen P. {et~al.}, 2020, Nature Methods, 17, 261

\end{thebibliography}

\label{lastpage}

\end{document}